\documentclass[%
 reprint,
 amsmath,amssymb,
 aps,
]{revtex4-1}

\usepackage{graphicx}
\usepackage{dcolumn}
\usepackage{bm}
\usepackage{amsthm, amsfonts, ifpdf, float, epstopdf}
\usepackage{caption}
\usepackage{subcaption}
\usepackage{multirow}
\usepackage{upquote}
\usepackage{color}

\begin{document}

\title{``Proxy-equation" paradigm - A novel strategy for massively-parallel asynchronous computations}

\author{Ankita Mittal}
 \email{ankitami@tamu.edu}
\author{Sharath Girimaji}
 \email{girimaji@tamu.edu}
\affiliation{Department of Aerospace Engineering, Texas A\&M University, College Station, Texas - 77843}

\begin{abstract}
Massively parallel simulations of transport equation systems call for a paradigm change in algorithm development to achieve efficient scalability. Traditional approaches require time synchronization of processing elements (PEs) which severely restricts scalability. Relaxing synchronization requirement introduces error and slows down convergence. In this paper, we propose and develop a novel `proxy-equation' concept for a general transport equation that (i) tolerates asynchrony with manageable added error, (ii) preserves convergence order and (iii) scales efficiently on massively parallel machines. The central idea is to modify \textit{a priori} the transport equation at the PE boundaries to offset asynchrony errors. Proof-of-concept computations are performed using a one-dimensional advection-diffusion equation. The results demonstrate the promise and advantages of the present strategy.
\end{abstract}

\maketitle

Massively parallel computing capability has the potential to reduce the total computational time of simulating large-sized complex physical systems. However, computing time-evolution of transport equations represents a special challenge. Such systems require overhead communication for synchronization among processing elements (PEs). Although advances in modern hardware and software have made it possible to communicate asynchronously \cite{Hardware}, mathematical level global synchronization is still a requirement for current numerical schemes. This imposed synchronization increases the idle wait time of PEs. Hardware and software failure rates increase with increasing number of PEs \cite{Failure} further adding to the PE load imbalance. Thus, the requirement of global synchronization throughout the computational domain leads to poor scaling characteristics with increasing system size \cite{Shriram, Petascale}. These imbalances become especially critical at exascale computing where millions of cores are expected to operate synchronously \cite{Exascale}. Therefore, it is important to develop computational strategies that tolerate asynchrony among PEs. 

\begin{figure}
\centering
  \includegraphics[width=.95\linewidth]{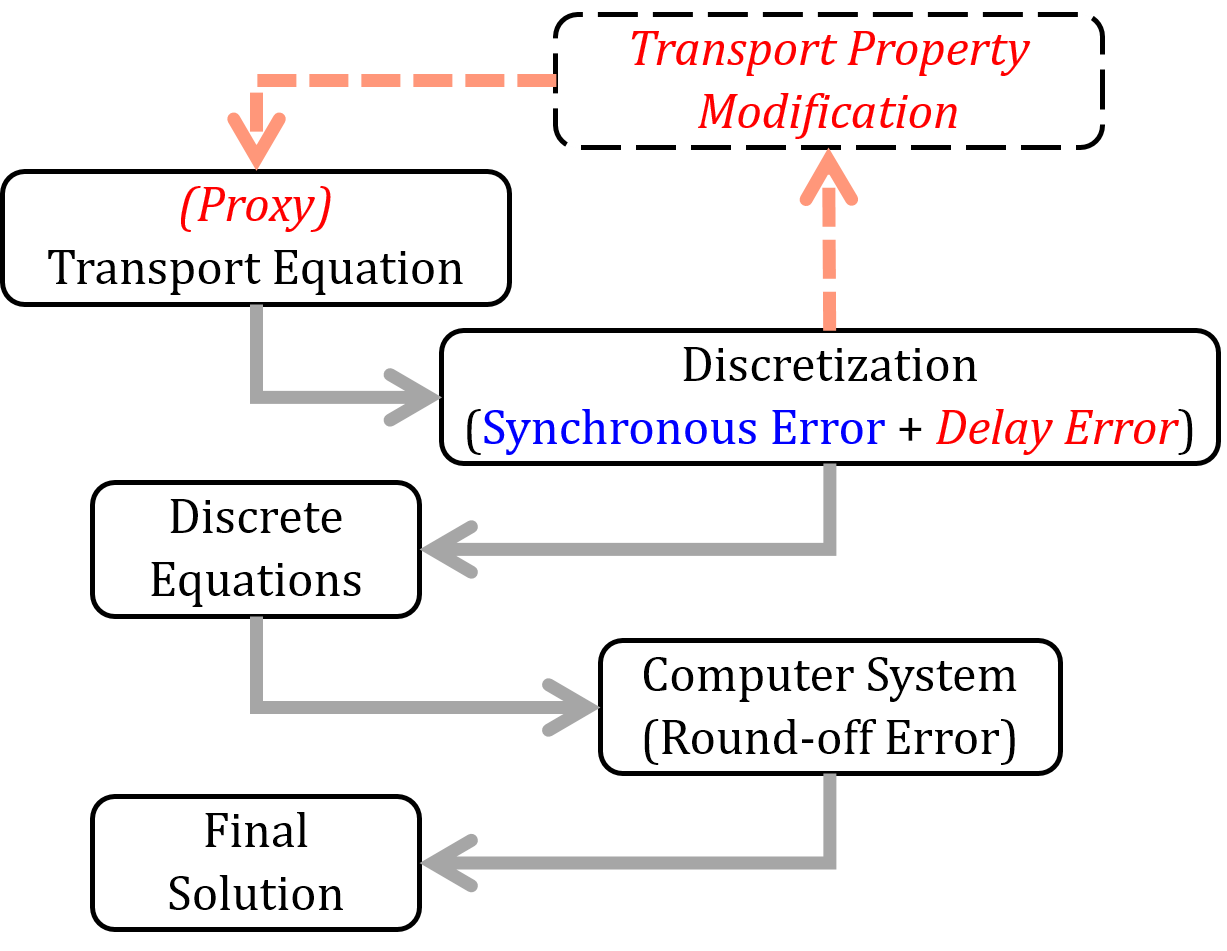}
\caption{Schematic of synchronous and asynchronous (in italics) errors. Logic of proxy-equation is also shown.}
	\label{fig:scheme_process}
\end{figure}

Typical synchronous computations (traditional methods) of transport equations incur truncation and round-off error. Asynchronous computations which relax the mathematical level synchronization develop additional error due to the  delay at PE boundaries called the delay error, FIG. \ref{fig:scheme_process}. Currently a few numerical schemes have been developed to improve the accuracy of asynchronous computations. Recently developed asynchrony-tolerant numerical scheme \cite{Aditya} and the delayed difference scheme \cite{PRL} attempt to counteract the delay error due to asynchrony by modifying the discretization scheme. However, the delay error still continues to be significant. Moreover, these approaches present difficulty when extending existing solvers to asynchronous computations. Modified numerical schemes also increase the stencil size which adds to the communication time.

In this paper, we present an alternate approach to mitigate the effect of asynchrony.  We start with the tenet that the function of asynchrony-tolerant computational strategy is to render the delay error to be of a lower order than the synchronous discretization error. To accomplish this, we propose modifying the governing equation as a function of delay, rather than changing the numerical scheme. The proposed modification is conceptually similar to the work of Warming \cite{Warming} and VonNeumann \cite{VonNeumann}. In that case, the modification was introduced to understand/improve robustness and stability of a synchronous scheme. In this paper, we develop the modified or {\em proxy} equation for the purpose of offsetting delay errors. The italicized (and red) parts of FIG. \ref{fig:scheme_process} identify the logic of the proxy-equation approach.

\textbf{\textit{Proxy-equation methodology:}} Advection-diffusion-reaction equation represents one of the most common transport systems in physics and engineering. Of these three effects, advection and diffusion include spatio-temporal communication and reaction is typically a local process. Thus, in this paper we will restrict ourselves to advection and diffusion phenomena as described by:
\begin{equation}
\frac{\partial u_i}{\partial t} + c_j \frac{\partial u_i}{\partial x_j} = \alpha \frac{\partial^2 u_i}{\partial x_j \partial x_j}
\label{Eq:ADE_3D}
\end{equation}
where, `\(c_j\)' represents the wave speed in each direction and `\(\alpha\)' represents the diffusion coefficient or viscosity. We will first consider the linear case wherein \(c_j\) is constant and then 
proceed to the non-linear case.

We analytically characterize the effect of asynchrony on advection and diffusion processes in isolation by examining the one-dimensional wave and heat equations:
\begin{equation}
\begin{split}
\frac{\partial u}{\partial t} + c \frac{\partial u}{\partial x} = 0 & \hspace{2 mm} \text{leading to,} \hspace{2 mm} u(x,t) = A e^{\iota \lambda(x - ct)} \\ \frac{\partial u}{\partial t} = \alpha \frac{\partial^2 u}{\partial x^2} & \hspace{2 mm} \text{leading to,} \hspace{2 mm} u(x,t) = A e^{\iota \lambda x} e^{-\alpha \lambda^2 t}
\label{Eq:TE_HE}
\end{split}
\end{equation}
The exact solutions of wave and diffusion equations are indicated for the initial condition of \(u(x,0) = A e^{\iota \lambda x}\). Here `\(\lambda\)' is the wavenumber of the initial field. Now, let us examine the computational solution of the above equations wherein a time delay of \(\delta t\) is introduced. Under delay conditions, the solution has to be inferred from the time instance \(t-\delta t\). From the form of the analytical solution it is evident that the effect of the delay error can be completely offset, if the wave-speed \(c^*\) and diffusion coefficient \(\alpha^*\) are redefined to satisfy,
\begin{equation}
c t = c^*(t-\delta t) \hspace{10 mm} \alpha t = \alpha^* (t - \delta t)
\label{Eq:mod_ex}
\end{equation}
leading to,
\begin{equation}
c^* \equiv \frac{c}{1-D_{fac}} \hspace{10 mm} \alpha^* \equiv \frac{\alpha}{1-D_{fac}}
\label{Eq:mod}
\end{equation}
Here, \(D_{fac}\) is the delay correction factor. Then the proxy-equation that can offset the delay effect can be written as,
\begin{equation}
\frac{\partial u}{\partial t} + c^* \frac{\partial u}{\partial x} = 0 \hspace{3mm} \text{and,} \hspace{3mm} \frac{\partial u}{\partial t} = \alpha^* \frac{\partial^2 u}{\partial x^2}
\label{Eq:px_eq}
\end{equation}
Clearly, the correction is a function of the degree of delay. This simple analysis demonstrates the manner of modification needed to mitigate advection and diffusion errors independently.
We now proceed to derive the delay correction factor for coupled computations.

\textbf{\textit{Correction factor determination:}} The exact delay correction factor, \(D_{fac}\) is determined by performing a truncation error analysis. Since the communication between PEs depend on the numerical scheme used, \(D_{fac}\) will vary with the scheme used as well as other factors such as the problem parameters, degree of delay and grid size. To illustrate the process, we proceed with a one-dimensional advection-diffusion equation, 
\begin{equation}
\frac{\partial u}{\partial t} + c \frac{\partial u}{\partial x} = \alpha \frac{\partial^2 u}{\partial x^2}
\label{Eq:ADE_1D}
\end{equation}
using a basic forward in time and central in space (FTCS) scheme. The synchronous stencil is,
\begin{equation}
\frac{u_i^{n+1}-u_i^{n}}{\Delta t} + c \frac{u_{i+1}^n - u_{i-1}^n}{2\Delta x} = \alpha \frac{u_{i+1}^n-2u_i^n+u_{i-1}^n}{\Delta x^2} + E_i
\label{Eq:Sync_scheme}
\end{equation}
where, \(E_i\) is the truncation error. To examine asynchronous effects we use the analytical approach as established in Donzis and Aditya \cite{Aditya}. Consider two processors PE0 and PE1 wherein PE1 is delayed by \(\tilde{k}\) timesteps in comparison to PE0 (right processor delayed). The advection-diffusion operator now has the form,
\begin{equation}
\frac{u_i^{n+1}-u_i^{n}}{\Delta t} + c \frac{u_{i+1}^{n-\tilde{k}} - u_{i-1}^n}{2\Delta x} = \alpha \frac{u_{i+1}^{n-\tilde{k}}-2u_i^n+u_{i-1}^n}{\Delta x^2} + E_i^{\tilde{k}}
\label{Eq:Async_scheme}
\end{equation}
where, \(E_{i}^{\tilde{k}}\) is the total error.
\begin{equation}
\begin{split}
E_{i}^{\tilde{k}} = & \frac{\Delta t}{2}\ddot{u} + \bigg(c \frac{\Delta x^2}{6} u^{'''}  - \alpha \frac{\Delta x^2}{12}u^{''''}\bigg) + \\ & \bigg(\tilde{k}\frac{\alpha \Delta t}{\Delta x^2} - \tilde{k}\frac{c \Delta t}{2\Delta x}\bigg) \dot{u} + ...
\label{Eq:Async_err}
\end{split}
\end{equation}
In Eq.(\ref{Eq:Async_err}), the first part of the error is due to truncation while the second part is the delay error. The delay error \(O(\Delta t/\Delta x^2)\) is of lower order than the truncation error \(o(\Delta x^2, \Delta t)\) and must be eliminated or reduced. The leading order terms in the delay error involve \(\dot{u}\) which also appears in the original equation. In order to reduce the delay error and improve accuracy, the original equation can be modified as,
\begin{equation}
\Bigg(1-\tilde{k}\bigg(r_{\alpha}-\frac{r_c}{2}\bigg)\Bigg)\frac{\partial u}{\partial t} + c \frac{\partial u}{\partial x} = \alpha \frac{\partial^2 u}{\partial x^2}
\label{Eq:Proxy_1}
\end{equation}
Eq.(\ref{Eq:Proxy_1}) presents one from of the proxy-equation for Eq.(\ref{Eq:ADE_1D}). This equation can be interpreted as either a system with added mass or modified time scale. The effects of asynchrony are pre-corrected by adding inertia to the system or by modifying the time scale near PE boundaries. The added mass coefficient or the delay correction factor for a FTCS scheme is given by,
\begin{equation}
D_{fac} \equiv \tilde{k}\bigg(r_{\alpha}-\frac{r_c}{2}\bigg), \hspace{3mm} r_{\alpha} = \frac{\alpha \Delta t}{\Delta x^2}, \hspace{3mm} r_c = \frac{c\Delta t}{\Delta x}
\end{equation}
Another representation of the proxy-equation can be obtained by dividing though by \((1-D_{fac})\) and can be written as, 
\begin{equation}
\frac{\partial u_i}{\partial t} + \frac{c_j}{(1-D_{fac})} \frac{\partial u_i}{\partial x_j} = \frac{\alpha}{(1-D_{fac})} \frac{\partial^2 u_i}{\partial x_j \partial x_j}
\label{Eq:ADE_Px}
\end{equation}
The Eq.(\ref{Eq:ADE_Px}) represents the form of the proxy-equation in line with the methodology discussed earlier. Here the transport parameters (i.e. advection speed and diffusion coefficient) are modified.

The above calculations are based on a right delay assumption. Performing a similar analysis for a left delay shows \(D_{fac} = \tilde{k}\Big(r_{\alpha}+\frac{r_{c}}{2}\Big)\). So the delay correction factor for a general 1D advection-diffusion equation solved using a FTCS scheme becomes,
\begin{equation}
D_{fac} \equiv \tilde{k}\Big(r_{\alpha}\pm\frac{r_{c}}{2}\Big)
\label{Eq:D_fac}
\end{equation}

For a three-dimensional transport equation Eq.(\ref{Eq:ADE_3D}), the delay correction factor can be found in a similar manner. For a FTCS scheme it is given by,
\begin{equation}
D_{fac} \equiv \sum_{r=1}^{d}\tilde{k}_r\Big(r_{\alpha,r}\pm\frac{r_{c,r}}{2}\Big), \hspace{3mm} r_{\alpha,r} = \frac{\alpha \Delta t}{\Delta x_r^2}, \hspace{3mm} r_{c,r} = \frac{c_r\Delta t}{\Delta x_r}
\label{Eq:D_fac_3D}
\end{equation}
where, `\(d\)' is the spacial dimension of the problem. Since this approach removes the leading order delay error, the order of accuracy of asynchronous computations improve by one. Higher order corrections will be considered in the future.

\textbf{\textit{Stability Analysis:}} Since a delay appears only at PE boundaries, performing an exact stability analysis is difficult. In order to determine a hard upper bound we consider the limiting case where delay appears at all grid points. We will focus on the stability of the proxy-equation under the \(\infty\)-norm \cite{Aditya}.

The modified transport coefficients in the proxy-equation are higher in magnitude compared to the original which leads to a restricted stability range. Performing some mathematical manipulation we can reduce the stability of the proxy-equation (for the FTCS scheme) to be as follows,
\begin{equation}
\frac{r_{c,r}}{2} \leq r_{\alpha,r} \hspace{2 mm} \text{and,} \hspace{2 mm}  \max_{r=1,...,d} r_{\alpha,r} \leq \frac{1}{2d[1+\max_{r} (\tilde{k}_r)]}
\label{Eq:stab_final}
\end{equation}
For no delay or \(\tilde{k}_r = 0\), we recover the original stability range. As the level of delay (\(\tilde{k}\)) increases, the stability range becomes more restricted which also sets a limit on the amount of delay that can be allowed. Other numerical approaches developed for asynchronous computations also present a similar issue with stability of the system. For stability, the upper bound in Eq.(\ref{Eq:stab_final}) is sufficient but not necessary. A true upper bound can only be determined numerically.

\begin{figure}
\centering
\begin{subfigure}{.5\textwidth}
  \centering
  \includegraphics[width=1.0\linewidth]{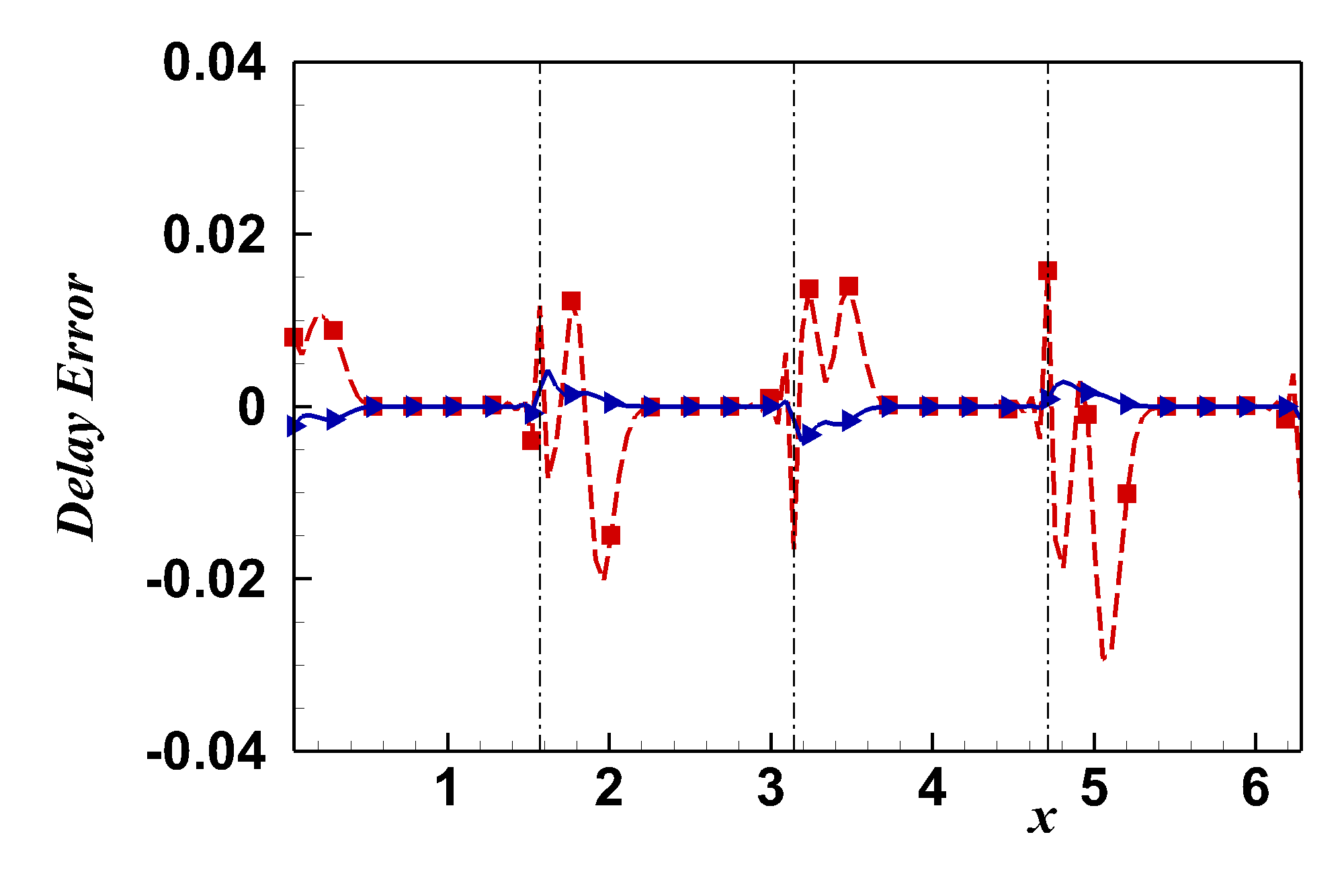}
  \caption{}
\end{subfigure}
\begin{subfigure}{.5\textwidth}
  \centering
  \includegraphics[width=1.0\linewidth]{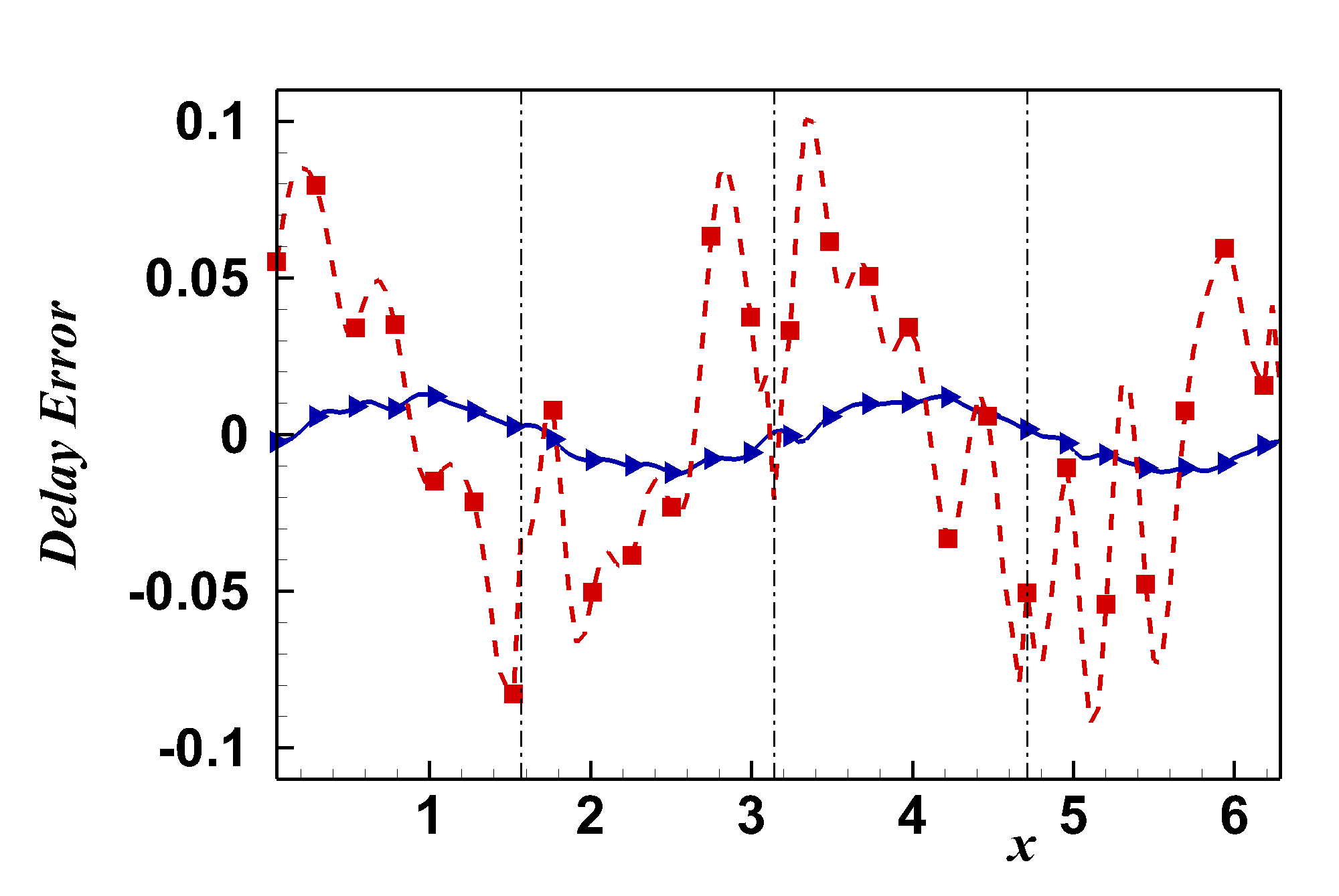}
  \caption{}
\end{subfigure}
\caption{The delay error, \(E_i^{\tilde{k}}-E_i\) for the original equation (red squares) and the proxy-equation (blue triangles) at (a) \(tc/l = 0.08\) (b) \(tc/l = 1.0\). Simulation parameters: N = 128, P = 4, \(c = 1\), \(\alpha = 0.01\), \(r_\alpha = 0.1\) with $\{$\(p_0,p_1\)$\}$ = $\{$\(0.3,0.7\)$\}$}
	\label{fig:ADE_Sol_Err}
\end{figure}
\textbf{\textit{Illustrations:}} We now present proof-of-concept numerical simulations using the one-dimensional advection-diffusion equation, Eq.(\ref{Eq:ADE_1D}) and its proxy-equation, Eq.(\ref{Eq:ADE_Px}) with the delay correction factor as in Eq.(\ref{Eq:D_fac}). Following Donzis and Aditya \cite{Aditya} a periodic domain of size \(l = 2\pi\) is considered with initial condition given by a collection of sinusoidal waves.
\begin{equation}
u(x,0) = \sum_k A(\lambda) \sin(\lambda x+\phi_k)
\label{Eq:initial}
\end{equation}
where, \(\lambda\) is the wavenumber with \(A(\lambda)\) and \(\phi_\lambda\) the amplitude and phase angle of each wave. The analytical solution to Eq.(\ref{Eq:ADE_1D}) with initial condition as in Eq.(\ref{Eq:initial}) is given by,
\begin{equation}
u_a(x,t) = \sum_\lambda e^{-\alpha \lambda^2 t} A(\lambda) \sin(\lambda(x-ct)+\phi_\lambda)
\label{Eq:Ex_Sol}
\end{equation}
We analyze the asynchronous effects using a similar computational approach as established in \cite{Aditya}. The results are presented for different grid sizes \(N = \{32, 128, 256, 512, 1024\}\), number of processors \(P = 4\) and two delay levels i.e. either delay of one time step or no delay at all. The delay is simulated in a manner similar to \cite{Aditya} using random number generators: \(p_0\) represents the probability of no delay and \(p_{1}\) represents the probability of one timestep delay (\(p_0 + p_1 = 1.0\)). \(p_0 = 1.0\) represents the synchronous case and \(p_0 = 0.0\) corresponds to the most asynchronous case. Results are presented for different values of \(p_0 = \{1.0,0.6,0.3,0.0\}\) in increasing order of asynchrony. Ensemble averages are taken over different initial phases for estimating order of error.
\begin{figure}
\centering
  \includegraphics[width=1.0\linewidth]{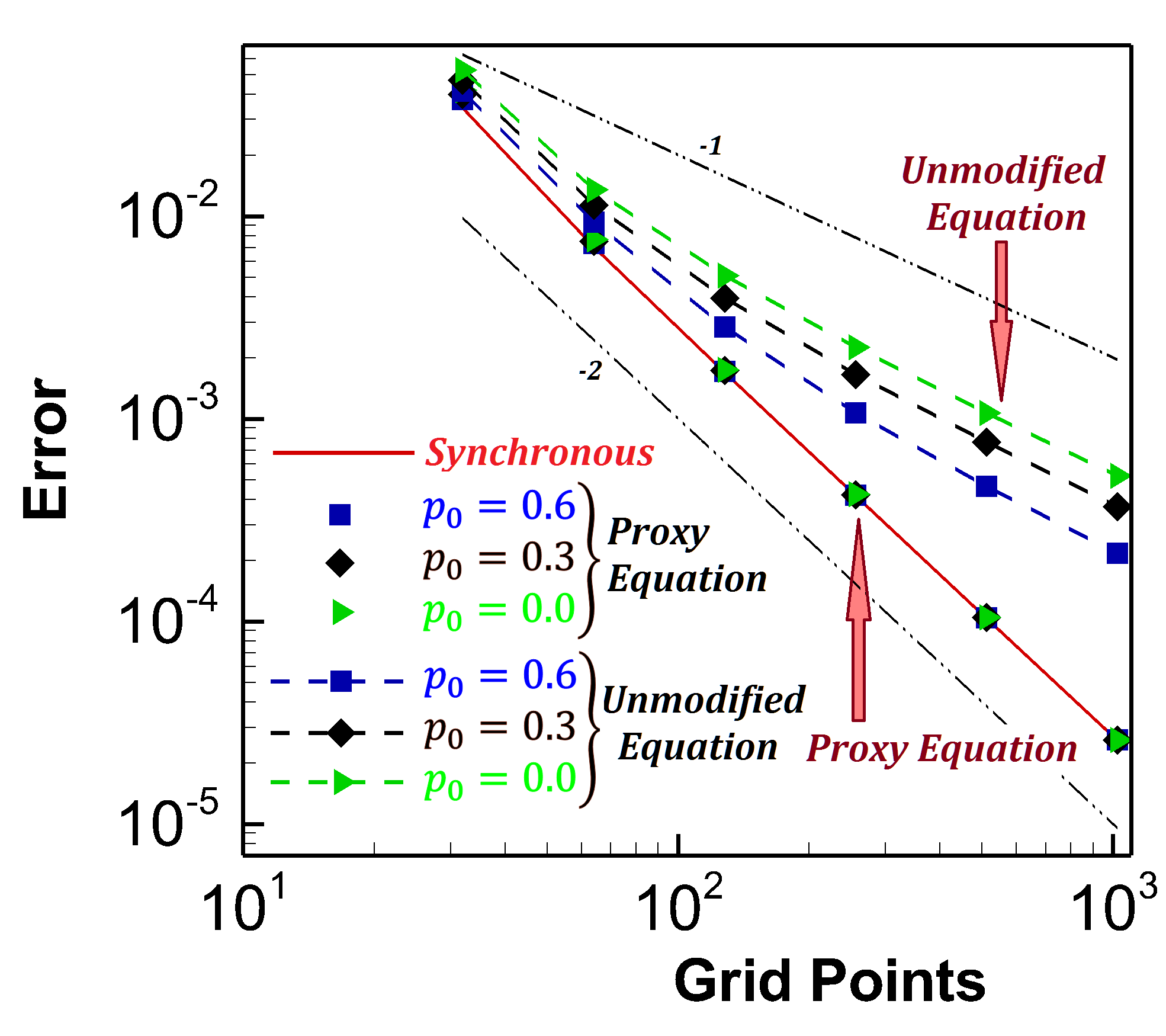}
\caption{Scaling of average error with number of grid points for proxy-equation and unmodified equation at \(tc/l = 1.0\) with \(r_\alpha = 0.1\), \(c = 1.0\), \(\lambda = 2\) and \(\alpha = 0.1\). All proxy-equation solutions fall on the same line.}
	\label{fig:ADE_Order}
\end{figure}

\begin{figure}
\centering
  \includegraphics[width=1.0\linewidth]{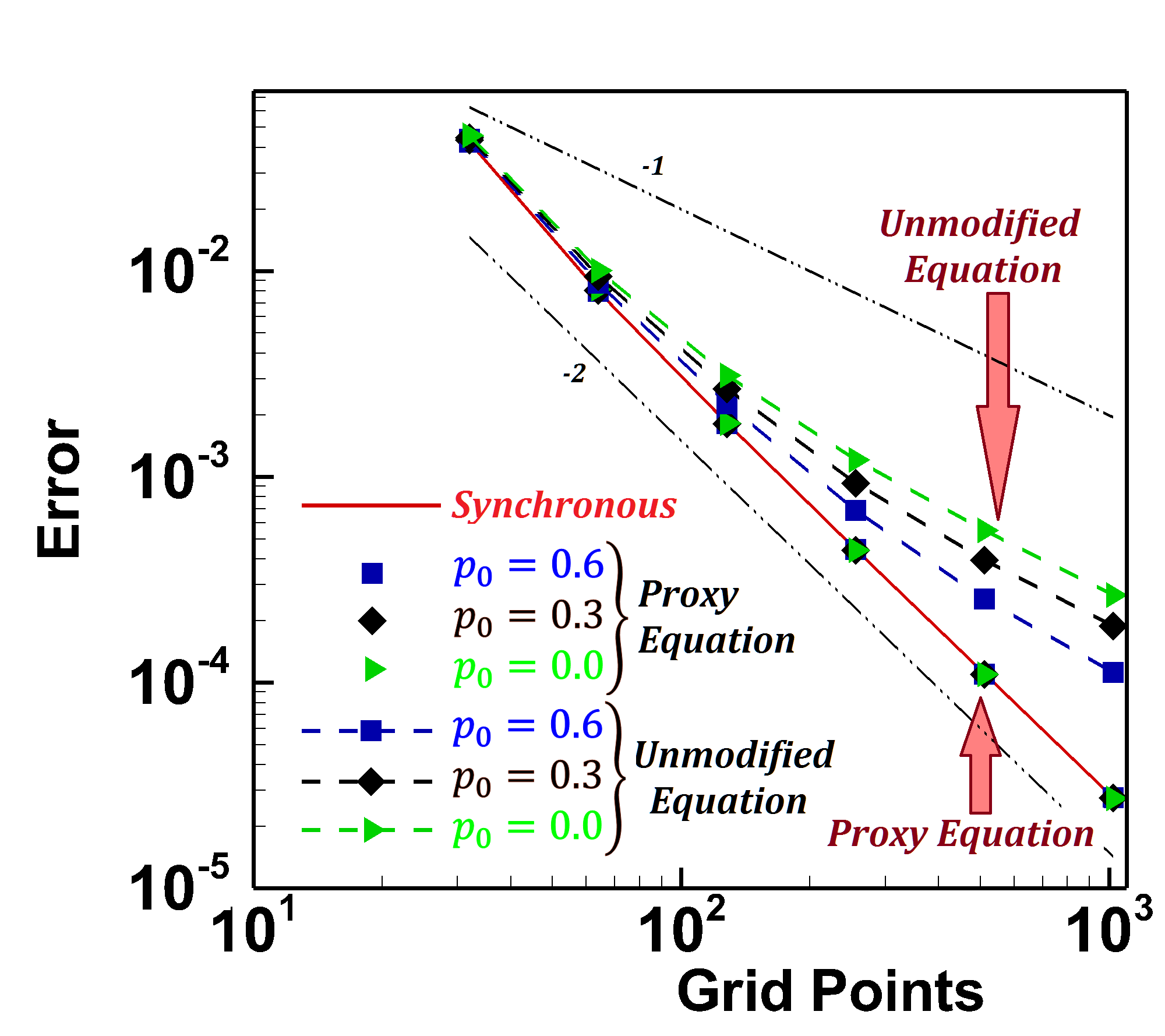}
\caption{Scaling of average error with number of grid points for proxy-equation and unmodified equation at \(tc/l = 1.0\) with \(r_\alpha = 0.1\), \(c = 1.0\), \(\lambda = 8\) and \(\alpha = 0.1\). All proxy-equation solutions fall on the same line.}
	\label{fig:ADE_Order_k8}
\end{figure}

\begin{figure}
\centering
\begin{subfigure}{.5\textwidth}
  \centering
  \includegraphics[width=1.0\linewidth]{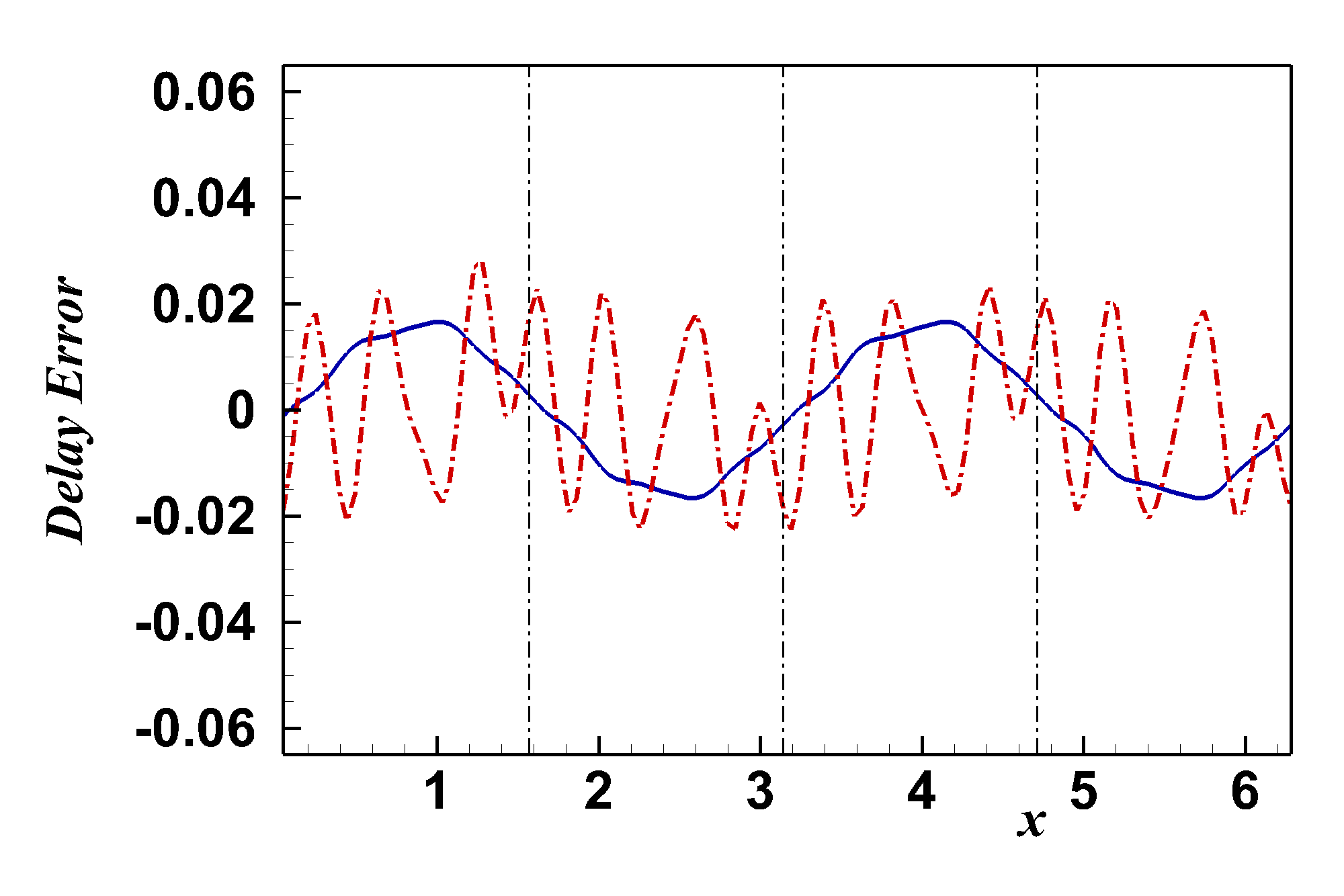}
  \caption{}
\end{subfigure}
\begin{subfigure}{.5\textwidth}
  \centering
  \includegraphics[width=1.0\linewidth]{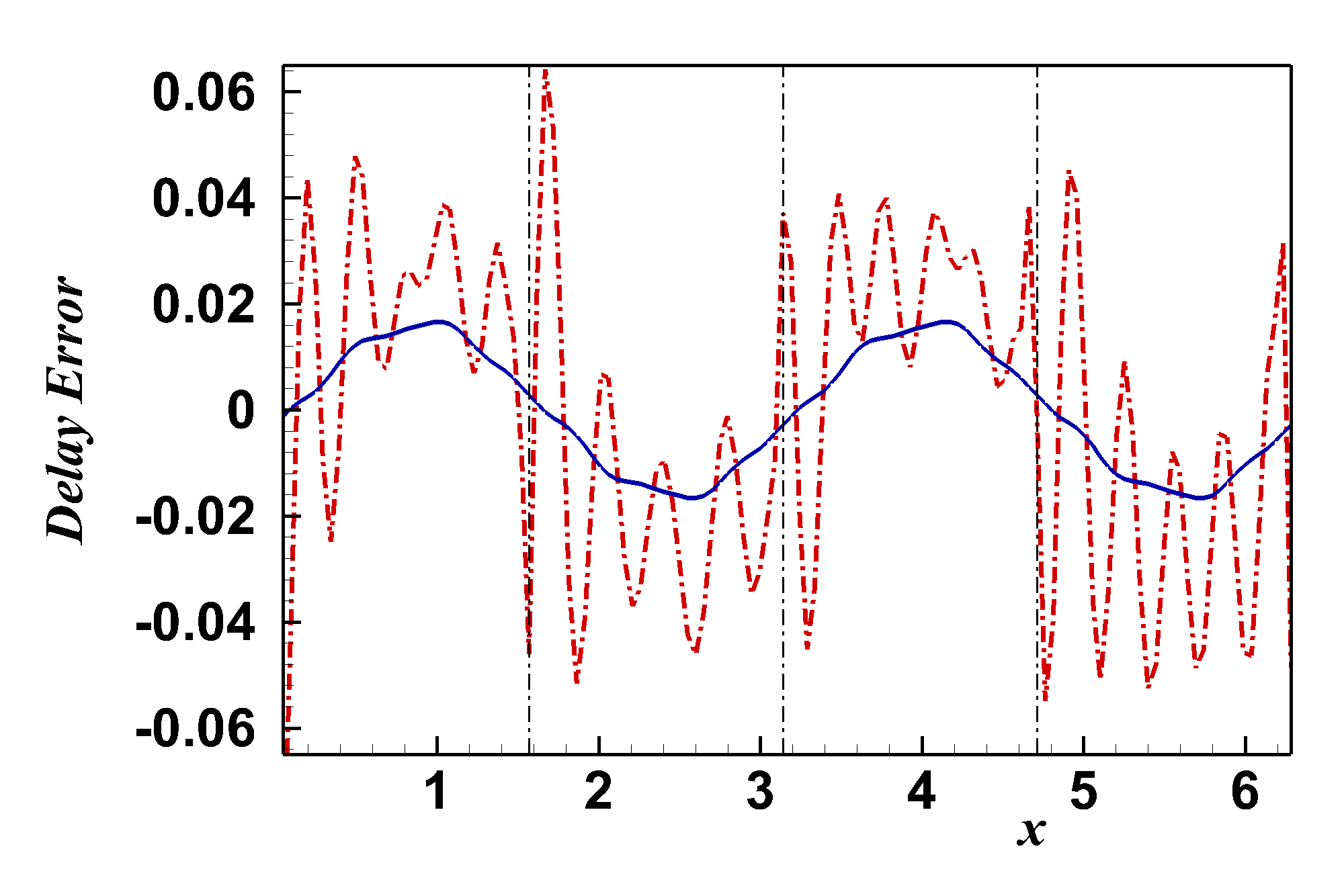}
  \caption{}
\end{subfigure}
\caption{Delay error comparisons at \(tc/l=1.0\) for the proxy-equation (solid blue line) and the original equation with (a) Ashynchrony tolerant scheme, (b) Delay difference scheme (dot dashed red). Simulation Parameters: N = 128, P = 4, \(c = 1.0\), \(\alpha = 0.01\), \(r_\alpha = 0.1\) with $\{$\(p_0,p_1\)$\}$ = $\{$\(0.3,0.7\)$\}$.}
	\label{fig:ADE_Err_comp}
\end{figure}

We present results for wavenumber, \(\lambda\) = 2. The plots in FIG. \ref{fig:ADE_Sol_Err} compare the delay error obtained when solving the original 1D advection-diffusion equation, Eq.(\ref{Eq:ADE_1D}) and its proxy-equation Eq.(\ref{Eq:ADE_Px}) under asynchronous computing using FTCS scheme at different times. The results clearly show the delay error initially manifests as a spike at the PE boundaries, slowly diffusing to interior grid points after a few time steps. At initial times (\(tc/l = 0.08\)) the delay error for the proxy-equation is much lower in magnitude, FIG. \ref{fig:ADE_Sol_Err}(a). Even when the simulation is run for considerably longer duration (\(tc/l = 1.0\)), the delay error for the proxy-equation stays considerably small and smooth, FIG. \ref{fig:ADE_Sol_Err}(b). In the present approach the delay error is neutralized as soon as it appears near the PE boundaries and thus, it does not diffuse to interior points at later time steps.

The scaling of average error (ensemble and space average) with increase in grid points (decreasing grid size) is shown in FIG. \ref{fig:ADE_Order}. The errors are evaluated at normalized time \(tc/l=1.0\). The synchronous case or \(p_0 = 1.0\) shows a second order scaling which is expected. Solving the original equation asynchronously without modification clearly shows a drop in the order of accuracy to first order. However, when the proxy-equation is used, the order of accuracy remains second order retaining the order of the original scheme. FIG.\ref{fig:ADE_Order_k8} shows the scaling of average error at a higher initial wavenumber (\(\lambda = 8\)). It clearly shows that even at higher wavenumbers the proxy-equation approach is valid and retains the order of accuracy for asynchronous computations.

\textit{Error Comparisons:} FIG. \ref{fig:ADE_Err_comp}(a) compares the delay error of the proxy-equation with the asynchrony-tolerant numerical scheme \cite{Aditya}. It clearly shows the delay error from the proxy-equation approach is much smother and also lower in magnitude. Similar observations can be made when comparing the delayed time difference scheme \cite{PRL} with the proxy-equation approach, FIG. \ref{fig:ADE_Err_comp}(b). Overall, it can be seen that the proxy-equation approach is significantly better at capturing the inherent physics of the problem.

\begin{figure}
\centering
  \includegraphics[width=1.0\linewidth]{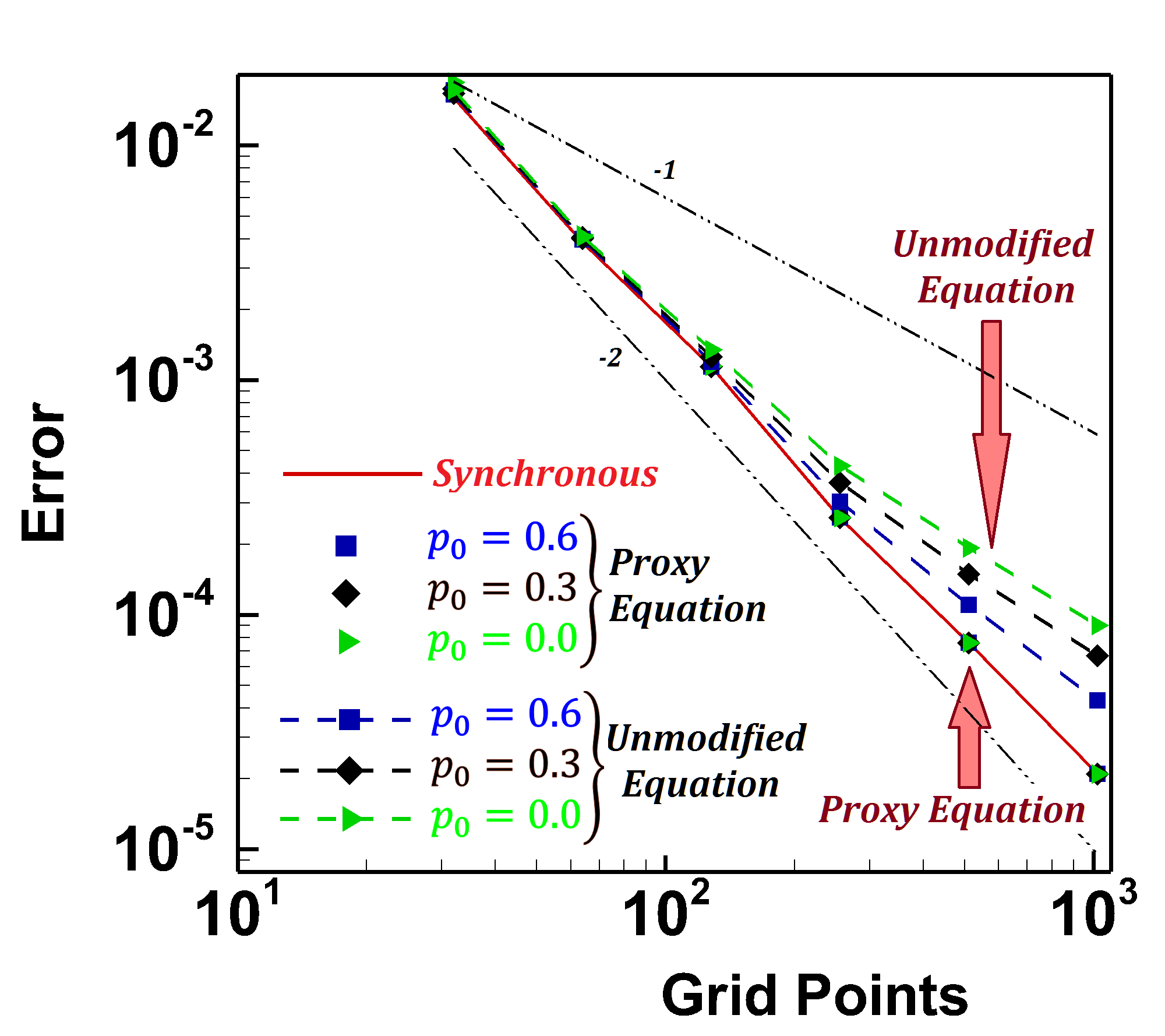}
\caption{Scaling of average error with number of grid points for proxy-equation and unmodified Burgers equation at \(tc/l = 1.0\) with \(r_\alpha = 0.1\) and \(\alpha = 0.1\). All proxy-equation solutions fall on the same line.}
	\label{fig:Bur_Order}
\end{figure}

\textit{Non-Linearity Effects:} Now we examine the effectiveness of the proxy-equation approach for a non-linear system using a one-dimensional viscous Burgers equation. 
\begin{equation}
\frac{\partial u}{\partial t} + u \frac{\partial u}{\partial x} = \alpha \frac{\partial^2 u}{\partial x^2}
\label{Eq:Bur}
\end{equation} 
The corresponding proxy-equation can be written as follows,

\begin{equation}
\begin{split}
(1-D_{fac})\frac{\partial u}{\partial t} + & u \frac{\partial u}{\partial x} = \alpha \frac{\partial^2 u}{\partial x^2} \\ D_{fac} = k\bigg(r_{\alpha} \pm \frac{r_u}{2}\bigg), \hspace{2 mm} & r_{\alpha} = \frac{\alpha \Delta t}{\Delta x^2}, \hspace{2 mm} r_u = \frac{u \Delta t}{\Delta x}
\label{Eq:Bur_proxy}
\end{split}
\end{equation}
FIG. \ref{fig:Bur_Order} shows average error scaling with grid resolution for the proxy-equation Eq.(\ref{Eq:Bur_proxy}) and the  Burgers equation Eq.(\ref{Eq:Bur}) under different levels of asynchrony. The results for the unmodified equation, Eq.(\ref{Eq:Bur}) again drop to first order while the ones for the proxy-equation, Eq.(\ref{Eq:Bur_proxy}) remain at second order as expected. The synchronous case is the red dashed line which is also second order.

In summary, we developed a modified equation or {\em proxy-equation} approach to mitigate the effects of asynchronous computations of transport equations on massively parallel computational systems.
A physical framework has been established to determine the degree of modification as a function of time delay between the processing elements (PEs). In principle, the proposed approach is similar to the
technique of \cite{Warming, VonNeumann} but with the intent of improving accuracy in the face of asynchronous computing. Various interpretations of the modifications have been established - a added mass system, time scale modification and modification of transport properties. The proxy-equation approach eliminates the need for any changes at the numerical scheme level making it easier to extend existing solvers to asynchronous computations. The advantages of the approach are demonstrated for both linear and non-linear advection-diffusion systems.

\end{document}